\documentclass[runningheads]{llncs}
\usepackage{graphicx}
\usepackage{multirow}
\usepackage{url}
\begin{document}
\pagestyle{empty}
\title{Structural subnetwork evolution across the life-span: rich-club, feeder, 
seeder}
\titlerunning{Structural subnetwork evolution across the lifespan}
%
\author{Markus D. 
Schirmer\inst{1,2,3\thanks{Corresponding author: mschirmer1@mgh.harvard.edu}}\and
Ai Wern Chung\inst{4}}
\authorrunning{Schirmer and Chung}
%
\institute{Stroke Division \& Massachusetts General Hospital, J. Philip Kistler 
Stroke Research Center, Harvard Medical School, Boston, MA, USA \and
Computer Science and Artificial Intelligence Lab, Massachusetts Institute of 
Technology, Cambridge, MA, USA \and
Department of Population Health Sciences, German Centre for Neurodegenerative 
Diseases (DZNE), Germany \and
Fetal-Neonatal Neuroimaging \& Developmental Science Center, Boston Children's 
Hospital, Harvard Medical School, Boston, MA, USA \\
\footnote{The final authenticated version is available online at \url{https://doi.org/10.1007/978-3-030-00755-3_15.}}
}

\maketitle
\begin{abstract}
The impact of developmental and aging processes on brain connectivity and the connectome has been widely studied. Network theoretical measures and certain topological principles are computed from the entire brain, however there is a need to separate and understand the underlying subnetworks which contribute towards these observed holistic connectomic alterations. One organizational principle is the rich-club - a core subnetwork of brain regions that are strongly connected, forming a high-cost, high-capacity backbone that is critical for effective communication in the network. Investigations primarily focus on its alterations with disease and age. Here, we present a systematic analysis of not only the rich-club, but also other subnetworks derived from this backbone - namely feeder and seeder subnetworks. Our analysis is applied to structural connectomes in a normal cohort from a large, publicly available lifespan study. We demonstrate changes in rich-club membership with age alongside a shift in importance from ‘peripheral' seeder to feeder subnetworks. Our results show a refinement within the rich-club structure (increase in transitivity and betweenness centrality), as well as   increased efficiency in the feeder subnetwork and decreased measures of network integration and segregation in the seeder subnetwork. These results demonstrate the different developmental patterns when analyzing the connectome stratified according to its rich-club and the potential of utilizing this subnetwork analysis to reveal the evolution of brain architectural alterations across the life-span.
\end{abstract}

\section{Introduction}
The last few decades have seen a rapid expansion in the application of network theory to study brain connectivity as it allows for a simple representation of complex systems~\cite{sporns2005human,bullmore2009complex}. Regions of the brain represent nodes in the network and edges can be defined either structurally or functionally, e.g. by using diffusion MRI (dMRI) or functional MRI (fMRI), respectively. These configurations have been widely studied across individual epochs of development, furthering our understanding of the development and function of the human connectome~\cite{sporns2004organization,sporns2010networks,tymofiyeva2012towards,schirmer2015developing,batalle2018annual}, as well as disease~\cite{collin2013impaired,ray2014structural,daianu2015rich}. 

The principle that the human connectome is divided up into subnetworks stems from the idea of functional segregation~\cite{bassett2006adaptive,sporns2013network}. To investigate differences in such subnetworks, studies may utilize a priori divisions of the connectome which are specifically targeted to their research question at hand~\cite{bonilha2013presurgical} or determine inter-connections exhibiting a statistical contrast through cluster-based thresholding~\cite{zalesky_network-based_2010}. However, studies for which such a division cannot be made a priori, data driven measures of modularity may be used, which fraction a connectome into modules~\cite{moussa2012consistency,cao2014topological,ingalhalikar2014sex}. Others utilize topological features, such as the structural core, to define a subdivision~\cite{hagmann2008mapping}. Whilst the connectome can be interrogated as a whole, identification and analyses of its subnetworks may reveal features with greater regional or network topological specificity and server as markers of change over time with disease or age.

Many topological aspects of the human brain have been studied, such as small-worldness~\cite{watts1998collective} and economically optimized wiring~\cite{bullmore2012economy}. Recently, another organizational principle was proposed - the rich-club (RC). The RC is a subset of nodes that is  more densely interconnected than expected by chance and has been studied in healthy controls~\cite{van2011rich}, during development~\cite{ball2014rich} and in disease~\cite{collin2013impaired,ray2014structural,daianu2015rich}. Moreover, this definition enables categorization of the edges by their association to the network's `backbone' that is its rich-club. This stratification identifies edges in the connectome as belonging to: RC, feeder (F) and seeder (S)~\cite{van2012high,ball2014rich,schirmer2015developing}. However, this classification has not been propagated to the nodal level to define subnetworks within a connectome, which can subsequently be studied across the life-span. 

In this work, we assess the change in properties of subnetworks, as defined by RC, F and S nodes, over the life-span. We first generate the nodal assignment to each category based on group-averaged connectomes for four age groups - younger than 20 years of age, between 20 and 40 years, between 40 and 60 years and above 60 years. We assess these group connectome assignments based on edge density and subsequently propagate the nodal labels back to each subject's connectome. Five subnetworks are defined: RC; F; S; RC and F combined (RC+F); and F and S combined (F+S). For completeness, we include the full connectome in the analyses. Finally, we calculate global network measures of betweenness centrality, efficiency, transitivity and assortativity for each subject and each subnetwork and each subject's connectome and investigate their association with age.

\section{Materials and Methods}
\subsection{Study design and patient population}
In this work we utilize data from the NKI/Rockland life-span study~\cite{nooner2012nki}. Preprocessed connectome data were obtained from the USC Multimodal Connectivity database\footnote[1]{http://umcd.humanconnectomeproject.org}~\cite{brown2012ucla}. MRI acquisition details are available elsewhere~\cite{brown2012ucla}. In brief, a total of 196 connectomes of healthy participants are computed, based on 3T dMRI acquisition (64 gradient directions; TR, 10000ms; TE, 91ms; voxel size, 2mm$^3$; \textit{b}-value, 1000 s/mm$^2$). Following eddy current and motion correction, diffusion tensors are modelled and deterministic tractography was performed, using fiber assignment by continuous tracking~\cite{mori1999three} (angular threshold $45^{\circ}$). Regions of interest (ROI) are based on the Craddock atlas~\cite{craddock2012whole}, resulting in 188 ROIs and connections are weighted by the number of streamlines connecting pairs of ROIs. Here, we normalize each connectome by the maximum streamline count for each subject, so that the connection weights $w_{ij}$ within each subject are $w_{ij} \in [0,1]$.

In our analysis, we divide the 196 participants into four age groups: Y20 $\leq$ 20 years, 20 years $<$ Y40 $\leq$ 40 years, 40 years $<$ Y60 $\leq$ 60 years, 60 years $<$ Y80 $\leq$ 80 years. Four subjects were above 80 years old (81, 82, 83 and 85 years). As there were only four subjects, we included them in the Y80 group. Table~\ref{tab:cohort} characterizes the study cohort and groups.

\begin{table}[ht!]
 \caption{NKI/Rockland life-span study cohort characterization and their division by age (in years).}
 \label{tab:cohort}
 \centering
 \begin{tabular}{|c|c|c|c|c|c|}
  \hline
  & Overall & Y20 & Y40 & Y60 & Y80 \\\hline
  N&196&53&67&47&29 \\\hline
  Age (mean (SD)) &35.0 (20.0)&13.41 (4.1)&27.4 (5.9)&47.4 (5.4)&71.0 (6.8) \\\hline
  Sex (Male; \%)&58.1&54.7&56.7&72.3&44.8 \\\hline
 \end{tabular}
\end{table}

\subsection{Group connectomes and rich-club organization}
A group-averaged connectome computed from weighted matrices is derived in two steps~\cite{van2011rich}. First, we calculate a binarized, group-average adjacency matrix by retaining edges that are present in at least 90\% of the subjects in each group. Weights are subsequently added to the group-averaged adjacency matrix by taking the average weight of each connection across the group, generating a weighted group-averaged connectome $W_{group}$ for each age group.

We utilize $W_{group}$ to subsequently calculate the weighted RC parameter $\phi_{group}(k)$~\cite{opsahl2008prominence} as implemented in the Brain Connectivity Toolbox~\cite{rubinov2010complex}, where k denotes the degree of nodes. The RC parameter $\phi_{group}(k)$ is normalized relative to a set of comparable random networks of equal size and with similar connectivity distribution. Here, we generate 100 random networks while preserving weight, degree and strength distributions of $W_{group}$~\cite{rubinov2010complex}. For each of these random realizations of the graph, we calculate the weighted RC parameter $\phi_{rand}(k)$. Finally, the normalized weighted RC parameter is calculated as

\begin{equation}
 \phi_{group}^{norm}(k) = \frac{\phi_{group}(k)}{mean(\phi_{rand}(k))}.
\end{equation}

For this metric, $\phi_{group}^{norm}(k)>$1 denotes the presence of a RC. We select 

\begin{equation}
 k_{max}^{group}: \phi_{group}^{norm}(k)>1,
\end{equation}

as the degree of the RC nodes of a given group, which allows us to determine the RC members. This analysis is repeated for each group and $k_{max}^{group}$ is recorded. We then use the mean of $k_{max}^{group}$ over all groups for all subsequent analyses.

\subsection{Subnetwork definition}
\begin{figure}[ht!]
 \centering
 \includegraphics[width=.8\linewidth]{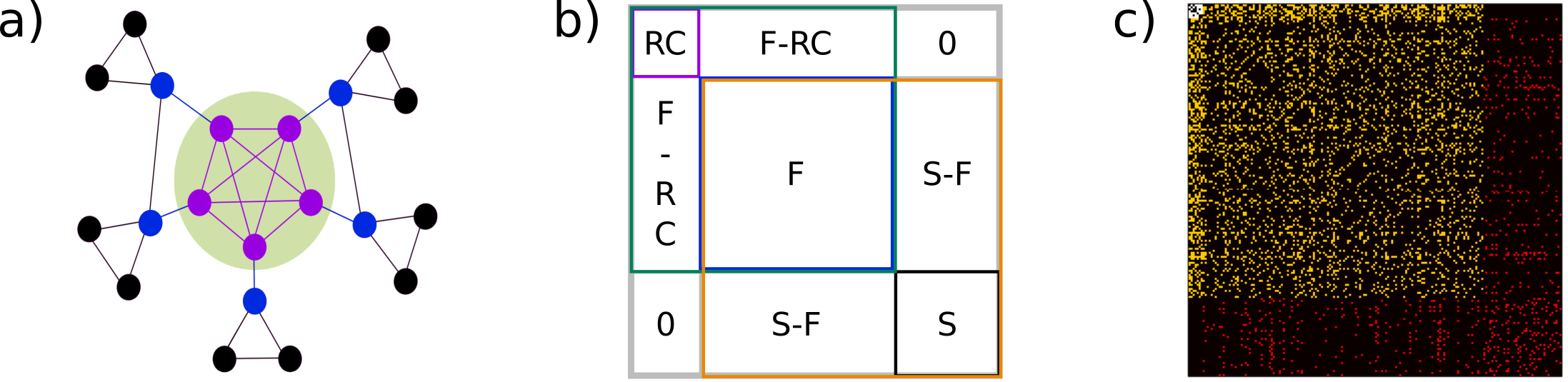}
 \caption{DDifferentiation of RC, F and S nodes. (a) Toy model of a network, showing the highly connected RC center (purple), F nodes (blue) with direct connections to the RC, but are not RC members themselves, and S
nodes (black) which are not RC members nor connected to the RC. (b) Regions within a connectivity matrix, classified by their membership to RC, F, or S nodes, as well as the connectivity between memberships (green, orange and gray). (c) Connectome of Y20 (≤20 years) group, rearranged to reflect the representation in b). Color-coded to represent RC (white), F (yellow) and S (red) connections.}
 \label{fig:figure1}
\end{figure}

In previous work, edges have often been differentiated after RC assessment, stratified according to their relation to the RC nodes - RC edges connect RC nodes; F edges connect RC to non-RC nodes; and S edges connect any two non-RC nodes~\cite{van2012high,ball2014rich,schirmer2015developing}. Similarly, nodes can be differentiated into RC, F and S, based on their connectivity profile. F nodes are directly connected to RC nodes, but do not belong to the RC themselves. S nodes do not belong to the RC nor are they directly connected to an RC node. Here, we define five subnetworks using this differentiation, namely RC alone (RC), F alone (F), S alone (S), as well as the subnetworks from combinations of RC, F and S, including the connections between them (RC+F; F+S; RC+F+S). These subnetworks are identified for each group using the group-averaged connectome and subsequently propagated to each subject. Figure~\ref{fig:figure1} illustrates a toy network example of and it's separation into RC, F, and S, with an example connectome for Y20. 

\subsection{Network analysis}
There are a variety of network measures which can be derived from any given graph or subnetwork, describing different local or global properties~\cite{watts1998collective,rubinov2010complex,van2011rich,chung2016characterising}. In this work we focus on global network measures of betweenness centrality (BC), global efficiency (E), transitivity (T) and assortativity (a). BC relates to the amount of information passing through a given node, thereby reflecting its importance for communication in the network. Although it is a local measure, the mean BC describes the prevalence of important nodes in a network. E characterizes how efficiently a network exchanges information. It also directly relates to the global integration of the network, where higher E reflects greater integration between specialized communities. T describes the likelihood of closed triangles in a network. If node n is connected to node m, which in turn is connected to node o, then T reflects the probability that node n is also directly connected to node o. These closed triangles lead to locally segregated networks, thereby promoting specialization~\cite{bassett2006adaptive,sporns2013network}. The last measure, a, is a correlation coefficient between the edge weights of node on opposite ends of a connection. If a is positive, it means that the connected nodes have a similar connectivity profile, reflecting a tendency for similar nodes in a network to be be linked. Importantly, these measures can be calculated in networks with multiple, disconnected components. After investigating the group differences of these four network measures, we calculate Spearman's correlation coefficient for each network measure against age for the entire NKI/Rockland cohort in each of the four groups to identify developmental trends. For the group-averaged connectomes, we further assess the edge density of the adjacency matrices for each of the five subnetworks.

\section{Results}
\subsection{Group connectome, rich-club and subnetwork definition}
Table~\ref{tab:table2} details the $k_{max}^{group}$ computed from each group-averaged connectome and the number of corresponding regions determined to form the rich-club subnetwork. The average $k_{max}^{group}$  over all groups quantifying the common degree for RC assessment is $k_{max}=55.25$. 

\begin{table}[ht!]
\centering
\caption{Normalized RC analysis of the four age groups, identifying the degree kmaxgroup, as well as the corresponding number of regions.}
\label{tab:table2}
 \begin{tabular}{|c|c|c|c|c|}
 \hline
  &Y20&Y40&Y60&Y80 \\ \hline
  $k_{max}^{group}$ & 56 & 54 & 55 &56 \\ \hline
  \# regions&8&11&10&7 \\ \hline
 \end{tabular}
\end{table}

\begin{figure}[ht!]
 \centering
 \includegraphics[width=.95\linewidth]{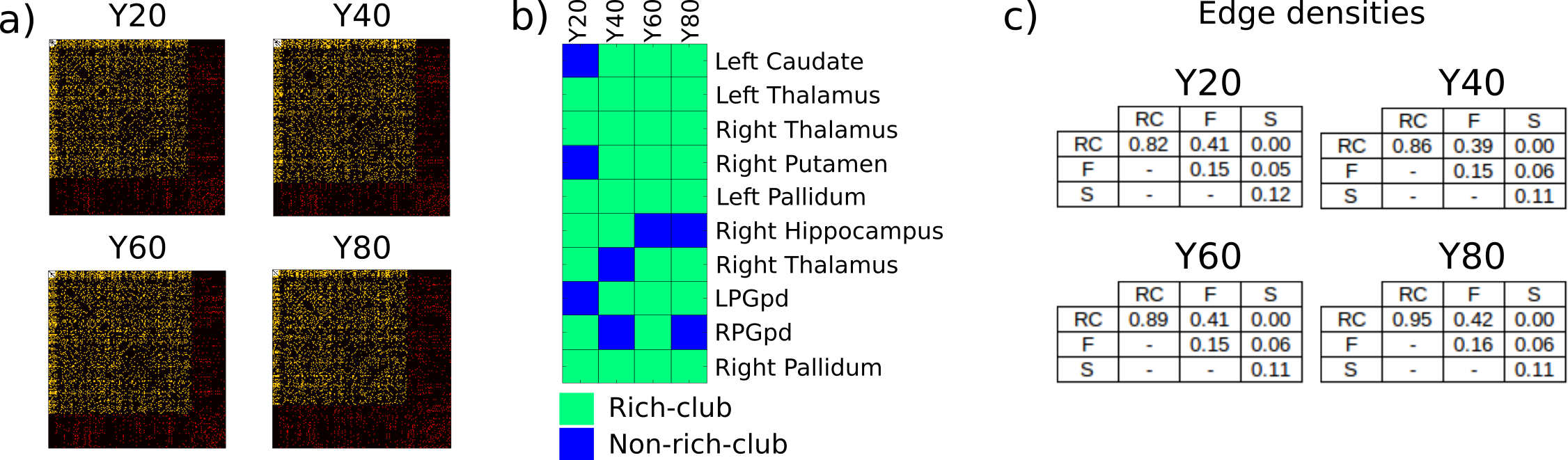}
 \caption{Membership analysis of the four age groups. a) Group-averaged connectomes for each group reordered as in Figure 1b, color-coded to represent RC (white), F (yellow) and S (red) connections. b) Brain regions identified as belonging to the RC for each of the four group-averaged connectomes (L/RPGpd: Left/Right Parahippocampal posterior). c) Edge density analysis in each section of the connectivity matrix (see Figure~\ref{fig:figure1}b) after nodal assignment to either RC, F, or S. }
 \label{fig:figure2}
\end{figure}

Using $k_{max}$ results in the identification of ten individual regions as belonging to the rich-club across the life-span cohort, with each region appearing in at least two of the four groups in our analysis (see Figure~\ref{fig:figure2}b). For each age group, Figure~\ref{fig:figure2}a shows the corresponding connectomes reordered by their relation to the rich-club, Figure~\ref{fig:figure2}b) lists the regions that are determined to be part of the RC for each group, and Figure~\ref{fig:figure2}c) shows the edge densities within and between nodal membership on the group-averaged connectomes.

\subsection{Subnetwork network analysis}
Following the RC, F, S differentiation, we label each subject's nodes with the corresponding assignment based on their respective group-averaged connectome. Figure 3 shows the distributions of mean values for each network theoretical measure computed from each subnetwork in each age group.

\begin{figure}[ht!]
 \centering
 \includegraphics[width=.95\linewidth]{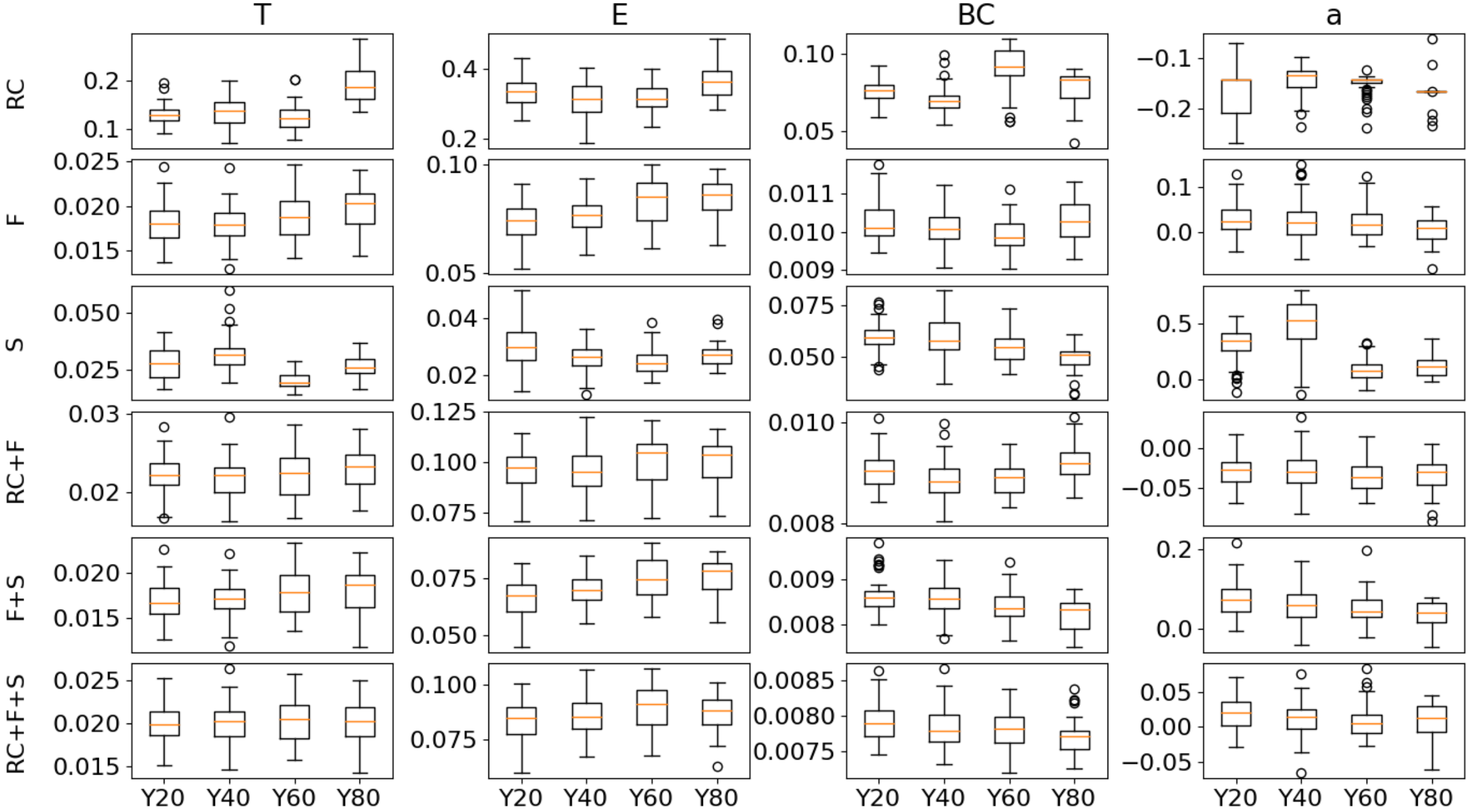}
 \caption{Network theoretical measures by subnetworks for each age group.}
 \label{fig:figure3}
\end{figure}

Utilizing the age of  all 196 subjects in the cohort, we calculate Spearman's rank correlation coefficient for each network measure within each subnetwork. Table~\ref{tab:table3} summarizes the developmental topological associations of each subnetwork with age.

\begin{table}[htbp]
\caption{Spearman's rank correlation coefficients between each network measure and age of the subjects for each subnetwork (*: p$<$0.05; **: p$<$0.01; ***: p$<$0.001).}
\begin{center}
\begin{tabular}{|c|c|c|c|c|c|}
\hline
 &  & \multicolumn{ 4}{c|}{Network measure} \\ \hline
 &  & T & E & BC & a \\ \hline
\multicolumn{ 1}{|c|}{} & RC & 0.26*** & 0.05 & 0.27*** & -0.10 \\ \cline{ 2- 6}
\multicolumn{ 1}{|c|}{\parbox[t]{2mm}{\multirow{3}{*}{\rotatebox[origin=c]{90}{Subnetwork}}}} & F & 0.21** & 0.37*** & -0.11 & -0.22** \\ \cline{ 2- 6}
\multicolumn{ 1}{|c|}{} & S & -0.29*** & -0.21** & -0.4*** & -0.46*** \\ \cline{ 2- 6}
\multicolumn{ 1}{|c|}{} & RC+F & 0.09 & 0.19** & 0.03 & -0.13 \\ \cline{ 2- 6}
\multicolumn{ 1}{|c|}{} & F+S & 0.18** & 0.37*** & -0.33*** & -0.34*** \\ \cline{ 2- 6}
\multicolumn{ 1}{|c|}{} & RC+F+S & 0.06 & 0.2** & -0.25*** & -0.2** \\ \hline
\end{tabular}
\end{center}
\label{tab:table3}
\end{table}

\section{Discussion}
In this work we presented a subnetwork analysis based on the differentiation of rich-club, feeder and seeder regions in the brain over the life-span. Utilizing group-averaged connectomes for subnetwork definition, we were able to identify developmental patterns in network measures from each subnetwork. Importantly, we demonstrated that rich-club, feeder and seeder regions evolve differently over time, potentially reflecting their functional difference in the human connectome. 

By dividing our cohort into individual groups based on their age, we showed that the determined degree for the presence of a  RC organization is largely consistent across age ranges. However, the number of regions identified as RC varied. We stabilized the number of RC regions by employing the average degree for our subnetwork analyses. In their study, van den Heuvel et al.~\cite{van2011rich} reported six bilateral regions (precuneus, superior frontal, superior parietal cortex, hippocampus, putamen and thalamus) making up the adult human connectome's RC. We note that our analyses comprised of only subcortical regions in the rich-club. While we were not able to reproduce the cortical RC regions found by van den Heuvel et al., our results agree with the remaining three subcortical RC regions (hippocampus, putamen and thalamus). In addition, we identified left caudate, and bilateral pallidum and parahippocampal posterior as part of the RC. 

Considering the variation across the age groups and the use of weighted connectomes, it is possible that RC membership may be a more fluid process, where specific regions may gain or lose their membership depending on age-specific requirements of the brain over the life-span (e.g. reflecting the neurobiological, metabolic and cognitive demands to meet developmental and aging processes). As such, others have similarly found this fluidity in rich-club membership in comparisons of children~\cite{grayson_structural_2014} and adolescents~\cite{dennis_development_2013} versus young adults. This is likely an effect of changing connectivity weights, as the edge density remains mostly constant for F, S, F-S and RC-F connections. It is only within the RC subnetwork that density increases with age. 

Analyzing network measures for each of the defined subnetworks demonstrates varying associations with age for each of the subnetworks. Our results show that the relationship between network measures with age is not linear, which is in agreement with other studies~\cite{zhao2015age,batalle2017early}. However, we observe clear trends for T and BC within the RC with age, suggesting that the RC may be further reinforced (increase in T) and that members of the RC might not be of equal importance in terms of information transport (increasing BC). Feeder regions are further integrated in the network, increasing the E of their connectivity with age, while further strengthening their communication paths within this subnetwork. Seeder, however, seemingly reduce in importance with age, shown in a reduction of every network measure investigated with age. The reduced importance may be indicative of a brain reorganization or pruning-like process, where increasingly limited resources for the human connectome are invested in the feeder regions.

There are limitations to our study. The identification of RC regions, though relatively stable across age groups, varies. This may be due to underlying biological processes, however, further investigation is necessary, in order to ensure that this is not the result of noise in the data or due to the cross-sectional nature of the study cohort. Furthermore, mostly subcortical regions were identified as being part of the RC. Future studies can use further divisions of the human connectome, e.g. into cortico-cortical and/or subcortico-subcortical connectomes, to elucidate more specific patterns of age-related change in the cortex. In addition, the use of multi-modal MRI data can further enhance our understanding of developmental and aging trajectories. Another important aspect will therefore be the use of combined structural and functional connectome data for analysis. 

In this study we report different patterns of evolution in subnetworks defined using rich-club topology over the life-span. We demonstrate in particular, that feeder regions are strengthened, while the seeder subnetwork is weakened with age. Future, multi-modal studies in healthy controls will allow the formulation of novel hypotheses, which can subsequently be tested in disease, and have the potential of identifying biomarkers for diagnoses and prognoses across a breadth of neuropathological and neurodevelopmental disorders. 

\section*{Funding}
This project has received funding from the European Union's Horizon 2020 research and innovation programme under the Marie Sklodowska-Curie grant agreement No 753896.  

%
%
%
\bibliographystyle{splncs04}

\end{document}